\documentclass[draft]{agujournal2019}
\usepackage{url} 
\usepackage[inline]{trackchanges} 
\usepackage{soul}
\usepackage{amsmath}
\usepackage{amssymb}
\draftfalse

\journalname{Journal of Advances in Modeling Earth Systems}

\begin{document}

%
%


\title{An Open-Source, Physics-Based, Tropical Cyclone Downscaling Model with Intensity-Dependent Steering}

%
%




\authors{Jonathan Lin\affil{1}, Raphael Rousseau-Rizzi{\affil{2}}, Chia-Ying Lee{\affil{1}}, Adam Sobel{\affil{1, 3}}}


\affiliation{1}{Lamont-Doherty Earth Observatory, Columbia University, Palisades, NY, USA}
\affiliation{2}{Institute of Geography, University of Bern, Bern, Switzerland}
\affiliation{3}{Department of Applied Physics and Applied Mathematics, Columbia University, New York, NY, USA}




\correspondingauthor{Jonathan Lin}{jlin@ldeo.columbia.edu}



\begin{keypoints}
\item The development of an open-source physics-based tropical cyclone downscaling model, based on random seeding, is described.
\item Steering of the tropical cyclone that is intensity-dependent can change tropical cyclone hazard on both local and regional scales.
\item The model reproduces the observed climatology of tropical cyclones, including the seasonal cycle, inter-annual variability, and hazard.
\end{keypoints}

%
%

%
%


\begin{abstract}
    An open-source, physics-based tropical cyclone downscaling model is developed, in order to generate a large climatology of tropical cyclones. The model is composed of three primary components: (1) a random seeding process that determines genesis, (2) an intensity-dependent beta-advection model that determines the track, and (3) a non-linear differential equation set that determines the intensification rate. The model is entirely forced by the large-scale environment. Downscaling ERA5 reanalysis data shows that the model is generally able to reproduce observed tropical cyclone climatology, such as the global seasonal cycle, genesis locations, track density, and lifetime maximum intensity distributions. Inter-annual variability in tropical cyclone count and power-dissipation is also well captured, on both basin-wide and global scales. Regional tropical cyclone hazard estimated by this model is also analyzed using return period maps and curves. In particular, the model is able to reasonably capture the observed return period curves of landfall intensity in various sub-basins around the globe. The incorporation of an intensity-dependent steering flow is shown to lead to regionally dependent changes in power dissipation and return periods. Advantages and disadvantages of this model, compared to other downscaling models, are also discussed.
\end{abstract}

\section*{Plain Language Summary}
Tropical cyclones are rare and extreme weather systems that can cause a lot of damage to society. Because the most intense of tropical cyclones are exceedingly rare, it is difficult to ascertain not only the frequency with which they occur, but how this frequency might change in the future. This problem is also compounded by the fact that even state-of-the-art climate models have trouble representing strong tropical cyclones. This study presents the development of a new, physics-based, tropical cyclone model. The model can rapidly simulate a large number of tropical cyclones given a mean climate, and is shown to reasonably reproduce the general behavior of tropical cyclones observed over the past 43 years. The model is open-source and freely available online.


%
%

%


%
%
%
%

\section{Introduction}
Tropical cyclones are extreme weather systems that are responsible for billions of dollars in damage to society every year \cite{pielke2008normalized}. As global warming continues, the consensus is that the frequency of intense tropical cyclones will increase \cite{knutson2010tropical, kossin2020global}. It follows that wind damage and precipitation will also increase with global warming \cite{emanuel2011global, knutson2020tropical}. Given the societal ramifications of tropical cyclones, it is prudent to understand not only tropical cyclone risk in the current climate, but also how the risk might change with warming.

Purely statistical models or statistical-dynamical models \cite{emanuel2006statistical} are often used to downscale tropical cyclone activity and estimate risk, instead of explicitly simulated tropical cyclones in reanalysis or climate models. One key reason for this is computational limitations on horizontal grid spacing in numerical models. Over a decade ago, various experiments showed that models with numerical mesh grid spacings of 50- to 260-km have severe negative biases in tropical cyclone intensity \cite{zhao2009simulations, hamill2011global, strachan2013investigating}. Advancements in computing power and numerical modeling led to studies showing that grid spacings of 10- to 25-km improve the models' ability to explicitly resolve the strong winds of TCs, though there is still difficulty representing the most intense of TCs \cite{davis2018resolving, magnusson2019ecmwf, roberts2020impact, roberts2020projected}. While these issues are generally remedied in global, convection permitting models \cite{judt2021tropical}, it is unclear whether convection permitting models will be ever able to be run on time scales long enough to robustly estimate tropical cyclone risk at regional and local scales. Even if numerical model resolution can be increased to eliminate negative biases in tropical cyclone intensity, the limitation on the length period with which these models can be run makes it extremely difficult to estimate the return period of the most intense TCs, which are often the ones that are of great societal interest. Thus, in general, tropical cyclone downscaling models have the desirable property of being able to rapidly simulate a large number of events given a certain climate, allowing for robust sampling of rare events. There is still much reason to develop, use, and understand tropical cyclone downscaling models.

In the recent decade, there have been a number of tropical cyclone downscaling models developed, though not all of them are open source \cite{lee2018environmentally, bloemendaal2020generation, jing2020environment, hong2020validation, xu2020design, chen2021typhoons}. All of these models have their own advantages and disadvantages, using a varying mixture of physics and statistics to generate a large number of synthetic tropical cyclones that are similar to historical tropical cyclones. In this paper, we describe the development of a publicly available, Python-based tropical cyclone downscaling model that synthesizes principles from the MIT tropical cyclone downscaling model \cite{emanuel2006statistical, emanuel2008hurricanes, emanuel2022tropical} and uses the FAST model to simulate tropical cyclone intensity given a large scale environment \cite{emanuel2017fast}. We have also incorporated a variety of changes to the downscaling model. In particular, we have expanded the FAST intensity model to the global scale, included an intensity-dependent steering level coefficient to the track model, introduced changes to the calculation of potential intensity to improve transparency, and incorporated a parameterization of the tropical cyclone ventilation that was previously evaluated in a tropical cyclone forecasting model \cite{lin2020forecasts}. The proposed model, available online at \url{https://github.com/linjonathan/tropical_cyclone_risk}, will help researchers in tropical cyclone climatology and risk to produce large datasets rapidly and transparently. The model is evaluated in the historical period by downscaling ERA5 reanalysis data \cite{hersbach2016era5}.

Section 2 describes the model in detail, including the genesis, track, and intensity algorithms. A thorough comparison with the observational record is shown in section 3. Section 4 explores tropical cyclone hazard on a global scale. Finally, section 5 concludes this study with a summary and discussion.

\section{Materials and Methods}

\subsection{Genesis}
This model uses random seeding, where seeds are randomly placed in space and time and allowed to evolve with the large-scale environment. This approach has been shown to successfully reproduce many aspects of tropical cyclone climatology \cite{emanuel2008hurricanes, emanuel2022tropical}. We also include a strong weighting function that depends on the background vorticity, similar to those used in genesis potential indices \cite{emanuel2004tropical, tippett2011poisson}. We use the function:

\begin{equation}
    P(\phi) = [(| \phi | - \phi_0) / 12]^\xi
\end{equation}
where $\phi$ is the latitude, $\phi_0$ is a tuning latitude parameter, and $\xi$ is a power-dependence that controls how quickly $P$ decays towards the equator. $P$ is not allowed to be smaller than zero or larger than unity. $P$ weights the random seeding, such that there are no seeds near the equator, where there are no observed tropical cyclones. Unlike the intensity model of \citeA{emanuel2004environmental}, the intensity model used in this study (described later) has no knowledge of the smallness of the Coriolis force near the equator. Without $P$ reducing the frequency of seeds near the equator, genesis will occur near the equator. In this model, we have chosen $\phi_0 = 2^\circ$. However, there is a some basin-to-basin variation in the optimal selection of $\xi$. This is important because $\xi$ partially controls the frequency of low-latitude genesis, which does exhibit some basin-to-basin variation in the observations. Thus, unlike genesis potential indices, which use a globally constant vorticity weighting function, we do vary $\xi$ by basin, as shown in Table \ref{table:parameters}. In each basin, $\xi$ was tuned such that the model's latitudinal distribution of tropical cyclone genesis matches that in the observations. This has the favorable effect of improving how well the genesis patterns, inter-annual variability, and return period curves compare to observations.

The seeds must also be initialized at a specific axisymmetric intensity, defined to be the azimuthal wind speed at the radius of maximum wind. An additional parameterization converts the axisymmetric wind speed to a maximum wind speed across the entire storm (which is the quantity reported in observations) and is described in the ensuing section. The seeds are initialized with an axisymmetric intensity of $v_{init}$, and only seeds that have intensified to at least $v_{2d} = 7$~m~s$^{-1}$ after 2 days, reach an axisymmetric intensity of at least $v_{min}$ and a maximum wind speed of at least $v^*_{min}$, are kept. As in \citeA{emanuel2022tropical}, the seeds must be initialized at a weak intensity to provide good statistics. Here, we use $v_{init} = 5$~m~s$^{-1}$. In order to accurately compare the downscaling model to observations, we use $v_{min} = 15$~m~s$^{-1}$ and $v^*_{min} = 18$~m~s$^{-1}$. Genesis is defined as when the seeds first reach a maximum intensity of $v^* = 18$~m~s$^{-1}$.

\subsection{Track Model}
After the seeds are initiated, they move in space and time according to the beta-and-advection model. The beta-and-advection model assumes that a tropical cyclone follows a weighted-average of the large-scale winds, plus a poleward and westward beta-drift correction that is a consequence of non-linear advection of the background vorticity gradient by the tropical cyclone winds \cite{marks1992beta}. Mathematically, this is:

\begin{equation}
    \textbf{v}_t = (1 - \alpha) \textbf{v}_{250} + \alpha \textbf{v}_{850} + \textbf{v}_{\beta} \cos(\phi)
    \label{eq_bam}
\end{equation}
where $\textbf{v}_t$ is the tropical cyclone translational vector, $\textbf{v}_{250}$ ($\textbf{v}_{850}$) is the large-scale environmental wind at 250-hPa (850-hPa), $\alpha$ is a steering coefficient, and $\textbf{v}_{\beta}$ is the translational speed correction due to beta-drift \cite{emanuel2006statistical}. In previous studies using this track model for tropical cyclone downscaling, a constant $\alpha = 0.8$ was chosen to minimize the 6-hour track displacement error from observations \cite{emanuel2006statistical}. Here, we iterate on this track model and provide evidence that the steering coefficient, $\alpha$, varies with intensity.

To show this, we evaluate the beta-and-advection model by finding what values of $\alpha$ minimize the error between the storm motion implied by the steering wind and that of the actual storm. To obtain the environmental steering wind, we performing ``vortex surgery" in reanalysis data, where the winds of the tropical cyclone are removed in order to calculate the background environmental steering winds within which each storm evolves \cite{galarneau2013diagnosing}. Since the divergence and vorticity of a tropical cyclone are typically elevated over its environment, the tropical cyclone's divergence and vorticity can be isolated from those of the environment and inverted, given suitable boundary conditions. The winds inferred from the inversion can then be subtracted from the full wind field to obtain the environmental wind. The reader is referred to \citeA{lin2020forecasts} for more details. We perform this vortex inversion on Atlantic, Eastern Pacific, and Western Pacific tropical cyclones from 2011-2021, using ERA5 reanalysis data over the same period. The tropical cyclones are identified using the International Best Track Archive for Climate Stewardship (IBTrACS) dataset \cite{knapp2010international}. Once the 250-hPa and 850-hPa environmental winds are obtained, we calculate the steering level coefficient $\alpha$ that maximizes the coefficient of determination, $r^2$, between the observed 6-hourly forward translational velocity and the translational speed predicted by Equation (\ref{eq_bam}), with $\textbf{v}_{\beta} = 0$. Note that $\textbf{v}_{\beta}$ is typically set to a constant. In this sense, $\textbf{v}_{\beta}$ has a larger influence on the mean-squared error and mean bias of the beta-and-advection model, and less so on $r^2$.

\begin{figure}
    \includegraphics[width=0.6\textwidth]{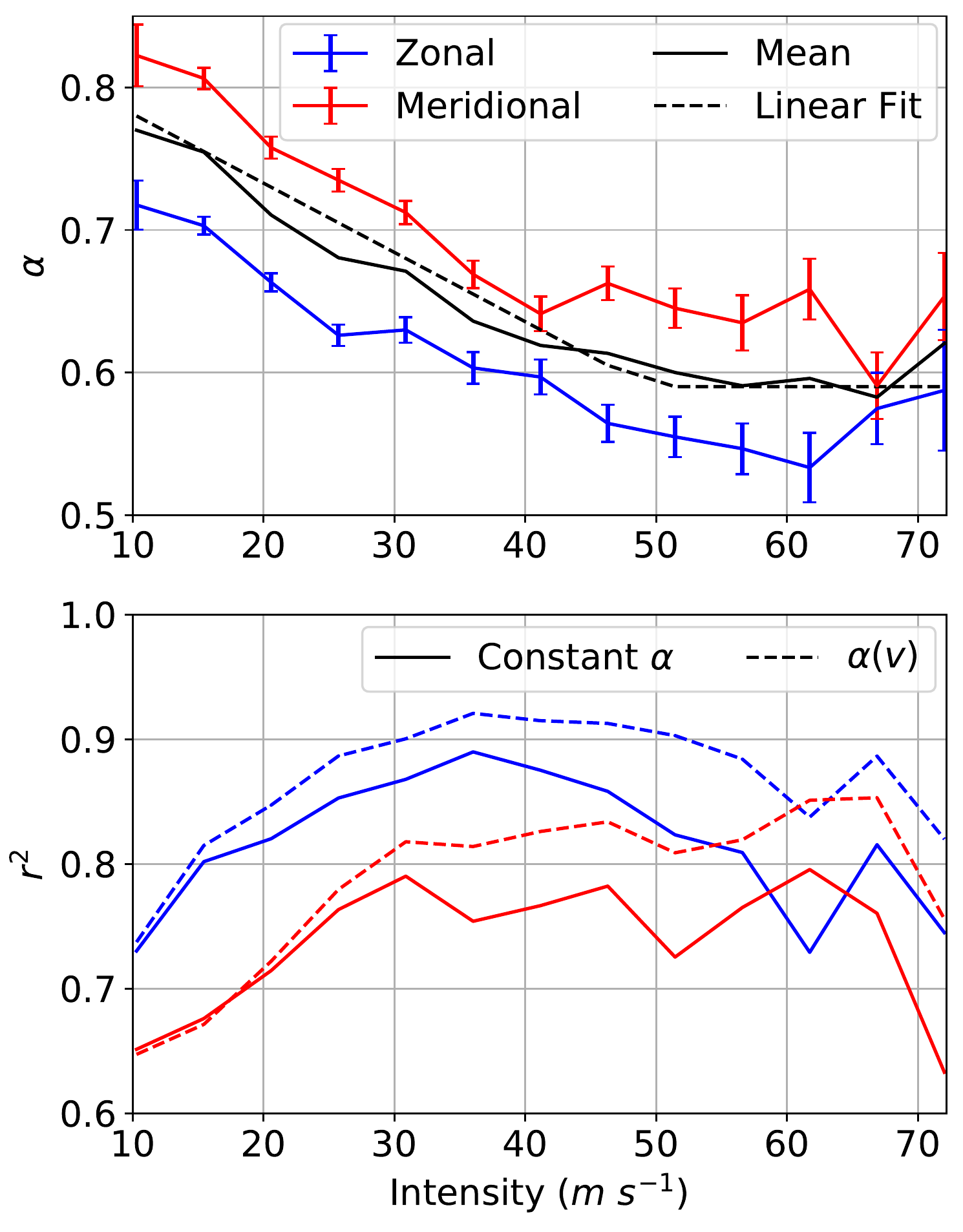}
    \caption{(Top): The steering level coefficient, $\alpha$ that maximizes $r^2$ between predicted and actual 6-hourly forward translational velocity in the (blue) zonal and (red) meridional directions. Error bars indicate the standard error. The solid black line is the mean between the zonal and meridional lines, while the dashed black line is the simple linear function to match the data. (Bottom): $r^2$ values between the predicted and actual forward translational velocity in the (blue) zonal and (red) meridional directions. Solid lines depict $r^2$ values using $\alpha = 0.8$, while dashed are for those using the simple linear function for intensity-dependent $\alpha$. Sample set includes Atlantic, Eastern Pacific, and Western Pacific tropical cyclones from 2011-2021, and the bin size is 5~m~s$^{-1}$, starting from 18~m~s$^{-1}$.}
    \label{fig:alpha_intensity}
\end{figure}

Figure \ref{fig:alpha_intensity}, top, shows the optimal $\alpha$ that maximizes $r^2$ for intensity bins of 5~m~s$^{-1}$ total width, starting from 10~m~s$^{-1}$. The optimal $\alpha$ decreases with intensity, but seems to level off to a constant after an intensity of 50~m~s$^{-1}$. This empirical relationship, which has also been qualitatively found in early studies of how the depth of the steering flow relates to tropical cyclone intensity \cite{dong1986relationship, velden1991basic}, indicates that the steering-level generally deepens as the tropical cyclone intensity increases. This is qualitatively consistent with the idea that as a tropical cyclone's circulation deepens, it is steered by winds further up in the atmosphere.

In light of this analysis, we introduce a simple linear function that describes the dependence of $\alpha$ on the intensity, $v^*$:

\begin{equation}
    \alpha(v_*) = \text{max} \{ \text{max} \{ v^* m_\alpha + b_\alpha,  \alpha_{\text{min}} \}, \alpha_{\text{max}} \}
    \label{eq_alpha}
\end{equation}
where $m_\alpha = 0.0013 \: \: \text{(m/s)}^{-1}$, $b_\alpha = 0.83$, set the slope and intercept of the linear function, and $\alpha_{\text{max}} = 0.78$, $\alpha_{\text{min}} = 0.59$ set the upper and lower bounds of $\alpha$. $\alpha$ is $\alpha_{\text{max}}$ at intensities weaker then 5~m~s$^{-1}$, and decreases linearly with increasing intensity until it is lower bounded by $\alpha_{\text{min}}$. The dependence of $\alpha$ on this empirical fit is shown in dashed-black in Figure \ref{fig:alpha_intensity}, top. Figure \ref{fig:alpha_intensity}, bottom, compares the $r^2$ of zonal and meridional translational velocities predicted by (solid) Equation \ref{eq_alpha} and (dashed) a constant $\alpha = 0.8$. The inclusion of a simple intensity dependent $\alpha$ leads to a significant increase in $r^2$ among all intensity bins. Furthermore, the mean-squared error of the translational velocity decreases for all intensity bins (not shown). The inclusion of the intensity-dependent $\alpha$ does not degrade or improve the mean statistics shown later in this paper. This is expected, as the majority of storms do not become major hurricanes. However, this finding is significant in the sense that it shifts the modeled spatial distribution of major tropical cyclone activity, as analyzed later in this study. It may also be important in the context of global warming, which is predicted to lead to an increase in tropical cyclone strength \cite{knutson2010tropical}, an expansion of the tropics \cite{seidel2008widening}, and increased poleward latitudes of tropical cyclone genesis \cite{sharmila2018recent}. An analysis of these potential effects with warming are left out of the scope of this paper, but will be investigated in future work.

Note that there is some variance in the slope of $\alpha$ with intensity by basin. It is not obvious why this is the case, but one potential source of uncertainty is the fact that linear interpolation between two levels, 250-hPa and 850-hPa, was used in determining the optimal $\alpha$. Some of the basin-to-basin variations could be explained through differences in the vertical structure of zonal and meridional environmental winds. However, the inclusion of more vertical levels in between 250- and 850-hPa is left to future work. 
As in \citeA{emanuel2006statistical}, stochastic realizations of the 250- and 850-hPa environmental winds are generated from monthly-averages and covariances of daily zonal and meridional winds at those levels. These stochastic realizations of the environmental wind are used to steer the seeded tropical cyclones according to Equation \ref{eq_bam} and \ref{eq_alpha}. Since we do not make any changes to the stochastic generation of environmental wind, the reader is referred to the supplement of \citeA{emanuel2006statistical} for more details.

\subsection{Intensity Model}
To evaluate the intensity of the tropical cyclone along the track, we use the FAST intensity model \cite{emanuel2017role, emanuel2017fast}, a simplified pair of coupled, non-linear ordinary differential equations that evolve $v$, the maximum azimuthal wind, and $m$, a non-dimensional inner-core moisture variable, given a particular environmental forcing. As stated in \citeA{emanuel2017fast}, $m$ can be thought of as a ``kind of relative humidity". The model equations are designed to reduce to the nonlinear analytical model of tropical cyclone intensification derived in \citeA{emanuel2012self}, under a fully water saturated inner core and zero environmental wind shear. This model was used successfully in a probabilistic tropical cyclone forecasting model \cite{lin2020forecasts}. 

The equations are included below for convenience:
\begin{eqnarray}
    \frac{dv}{dt} =& \frac{1}{2} \frac{C_k}{h} \Big[ \alpha \beta V_p^2 m^3 - (1 - \gamma m^3) v^2 \Big] \label{dvdt} \\
    \frac{dm}{dt} =& \frac{1}{2} \frac{C_k}{h} \Big[ (1 - m) v - \chi S m \Big] \\
    \beta =& 1 - \epsilon - \kappa \\
    \gamma =& \epsilon + \alpha \kappa \\
    \epsilon =& \frac{T_s - T_o}{T_s} \\
    \kappa =& \frac{\epsilon}{2} \frac{C_k}{C_d} \frac{L_v}{R_d} \frac{q_s^*}{T_s} \\
    \alpha =& 1 - 0.87 \exp^{-z} \\
    z =& 0.01 \Gamma^{-0.4} h_m u_T V_p v^{-1} \\
    \chi_{\text{grid}} =& \frac{s^* - s_m}{s^*_0 - s^*}
\end{eqnarray}
where $C_k$ and $C_d$ are the surface enthalpy and drag coefficients, $h$ is the atmospheric boundary layer depth, $V_p$ is the potential intensity, $\alpha$ is an ocean interaction parameter, $\chi_{\text{grid}}$ is the gridded mid-level saturation entropy deficit, $s^*$ ($s_0$) is the saturation moist entropy of the free troposphere (sea surface), $s_m$ is the moist entropy of the middle troposphere, $S$ is the 250-850-hPa vertical wind shear, $T_s$ is the surface temperature, $T_o$ is the outflow temperature, $L_v$ is a constant latent heat of vaporization, $R_d$ is the dry gas constant, $q^*_s$ is the surface saturation specific humidity, $\epsilon$ is the thermodynamic efficiency, $\alpha$ is an ocean interaction parameter, $\Gamma$ is the sub-mixed layer thermal stratification in $K \: (100 \: \: m)^{-1}$, $h_m$ is the mixed layer depth, and $u_T$ is the translation speed. The reader is referred to~\citeA{emanuel2017fast} for further details. For the purposes of simplicity, we take $\beta$, $\gamma$, $\epsilon$, and $\kappa$ to be constant. As such, the key environmental quantities that drive differences in the intensification of a tropical cyclone in this model are the potential intensity $V_p$, the vertical wind shear $S$, the environmental entropy deficit $\chi$, and the ocean interaction parameter $\alpha$. The vertical wind shear is taken from the synthetic realizations of the upper- and lower-level winds, while the ocean interaction parameter is evolved using climatological profiles of ocean mixed-layer depth and sub-mixed layer thermal stratification. It is possible that using reanalysis estimates of ocean mixed-layer depth and sub-mixed layer thermal stratification could lead to improvements of these results. There are some changes made to the calculations of potential intensity and environmental entropy deficit, which are outlined in the next sections.

Since the FAST equations are a coupled set of ordinary differential equations, both $v$ and $m$ need to be initialized. $v$ is given from the random seeding approach, and thus we are left with a choice on how to initialize $m$. Following \citeA{emanuel2017fast}, which initialized $m$ as a function of the large-scale relative humidity, we choose to initialize $m$ as a logistic curve of the large-scale monthly mean relative humidity.

\begin{equation}
    m_{\text{init}} = \frac{L}{1 + \exp(-k (\mathcal{H} - \mathcal{H}_0))} + m_0
\end{equation}
where $L = 0.20$, $k = 10$, $\mathcal{H}_0 = 0.55$, $m_0 = 0.125$, and $\mathcal{H}$ is the large-scale relative humidity. This equation was arrived at somewhat empirically, but with the general idea that a moister large-scale environment is more conducive to tropical cyclone genesis. Note that this is different from the initialization of $m_{\text{init}} = 1.2 \mathcal{H}$ chosen by \citeA{emanuel2017fast}, which leads to intensification rates much larger than observed. 

Finally, since the FAST equations predict only the axisymmetric wind, $v$, a conversion to the maximum wind speed $v^*$ (to easily compare with observations) must be performed. We follow the same model optimized in \citeA{lin2020forecasts}, adding a wind vector that is a function of the translational speed and large-scale environmental wind to convert $v$ into $v^*$. The equation is detailed in the Appendix.

\subsubsection{Potential Intensity}
\begin{figure}
    \includegraphics[width=0.75\textwidth]{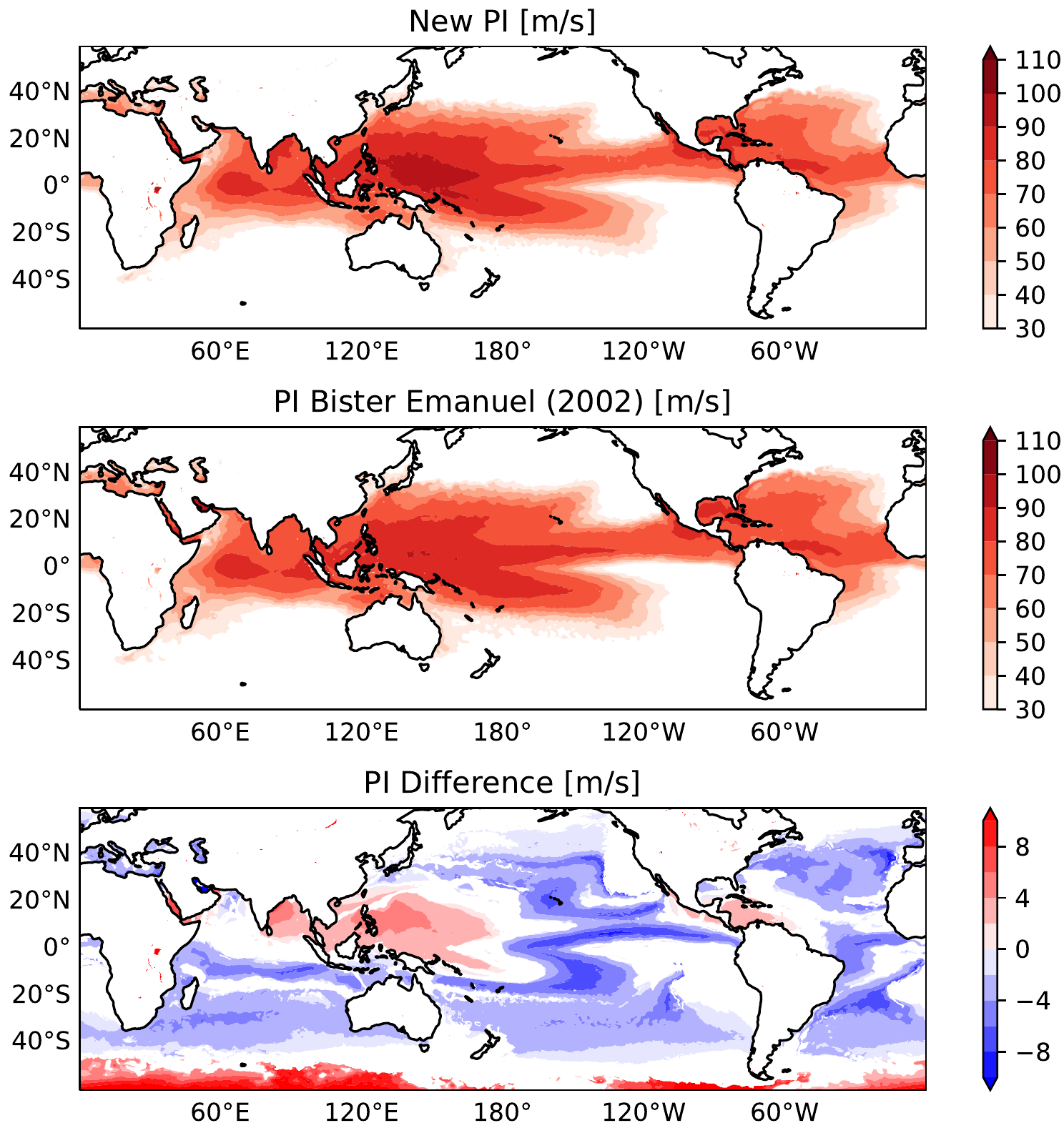}
    \caption{2017 boreal hurricane season (August-September-October) average PI, computed using (top) the new algorithm and (middle) the algorithm of BE02. (Bottom) Average of new PI minus that computed using BE02, for the 2017 boreal hurricane season.}
    \label{fig:PI_comp}
\end{figure}

Along with this Python-based model, we briefly describe a new Python-based algorithm for calculating potential intensity (PI, or $V_p$). This new algorithm is a version of the MATLAB algorithm introduced by \citeA[hereafter BE02]{bister2002low}, which was modified to run faster and be more modular and transparent. As in previous algorithms, $V_p$ is calculated from environmental soundings using the formula

\begin{equation}
    V_p^2 = S_w^2 \frac{C_k}{C_d} \frac{T_s}{T_o}(CAPE^* - CAPE),
\end{equation}

where $CAPE$ and $CAPE^*$ are respectively the environmental convective available potential energies of a near-surface parcel and of a surface saturated parcel at temperature $T_s$. $S_w$ is an empirical constant used to reduce PI-estimated wind-speeds to surface wind speeds observed in tropical cyclones. A value $S_w = 0.8$ is chosen, loosely based on the work of \citeA{powell1980evaluations}. In the CAPE computations, the lower condensation level is computed using the formula of \citeA{romps2017exact}. Model options include computing ascent profiles for CAPE using either pseudoadiabatic \cite{bryan2008computation} or reversible \cite{emanuel1994atmospheric} definitions of moist entropy. The new algorithm considers the effects of dissipative heating on storm intensity \cite{bister1998dissipative}, but not the effect of central pressure drop on eyewall enthalpy transfer \cite{emanuel1988maximum} considered in BE02. While one may argue that by neglecting the iterations on central pressure we are neglecting a physically important mechanism, we find no monotonically increasing difference between $V_p$ computed using our algorithm and $V_p$ computed using the algorithm of BE02 with identical $S_w$ and exchange coefficients. In addition, $V_p$ is not a quantity that can be observed, but instead must be estimated from environmental conditions using different algorithms or even formulas, all subject to different assumptions \cite{rousseau2022connection}. Hence, here we do not aim for a perfect correspondence between our PI algorithm and that of BE02, but one sufficient to warrant its use here. Results from the new PI algorithm and that of BE02 are compared in Figure \ref{fig:PI_comp} for the particularly active hurricane season of 2017. The figure shows that, qualitatively, the two algorithms produce very similar results, with the new algorithm producing somewhat lower PI in subsidence regions, subtropics and midlatitudes, and higher PI in strongly convecting regions of the deep-tropics. This result suggests that neglecting the effect of central pressure drop on enthalpy transfer in our algorithm is not a problem. If it were, the algorithm of BE02 should produce relatively higher values in the deep tropics, where PI is already high. The differences between the two algorithms are usually less than 5$\%$. A histogram further comparing the values of PI computed using the two algorithms is available in Supporting Information, Figure S1.

Computing CAPE requires inverting moist entropy to obtain parcel temperature profiles on pressure levels is a time consuming computational step. Here, we make use of the fact that the range of temperatures and pressures is not large up to the tropopause, and we provide the user with the option to pre-compute tables of temperature in entropy and pressure coordinates. In these tables, each combination of entropy and pressure needs to be inverted only once to obtain temperature. Then, the computationally costly step of inverting the entropy equation to compute CAPE becomes a simple interpolation. We find that interpolation using a pseudoadibatic entropy table with equally spaced pressure coordinates ranging from 25- to 1050-hPa and entropy coordinates ranging from 2300 to 3600 J Kg$^{-1}$ K$^{-1}$ yields negligible differences from inversion when the table resolution is at least 100$\times$100. Reversible entropy interpolation tables require an additional ``total water mixing ratio'' dimension. Note that computing these tables only requires inverting moist entropy between 1e4 and 1e6 times, while computing $V_p$ globally at a single time for a coarsely resolved (e.g., 2.5 degrees and 15 vertical levels) climate simulation requires inverting moist entropy about 3e6 times. \citeA{gilford2021pypi} estimated that the time required for computing $V_p$ at 1e5 points using the BE02 algorithm is 8.5s for the original Matlab implementation and 10s for their Python implementation. The new algorithm used here runs in 2 seconds for pseudoadiabatic and 3 seconds for reversible thermodynamics (this difference is due to the additional dimension of the reversible entropy interpolation table). In addition, the new algorithm is vectorized and designed to be run in parallel. Computing monthly-mean PI over 100 years of climate simulation at 1 degree resolution (8e7 points) and on 10 cores takes less than 3 minutes.

\subsubsection{Entropy Deficit}
The ventilation of the tropical cyclone, or drying of the inner-core \cite{tang2010midlevel}, is parameterized in the FAST system through the term $\chi S m$. In~\citeA{emanuel2017fast}, the entropy deficit is set as a constant, $\chi = 2.2$. However, the entropy deficit increases with warming, and has been shown to play a critical role in controlling the number of tropical cyclones predicted by downscaling models \cite{emanuel2013downscaling, lee2020statistical}, statistical indices extrapolated to future climates \cite{camargo2014testing}, and explicit numerical models \cite{hsieh2022model} under future warming.

In the probabilistic tropical cyclone forecasting model, variations of moisture on daily timescales are important in setting the spatial distribution of saturation entropy deficit. In \citeA{lin2020forecasts}, $\chi$ is parameterized as approximately the 90th percentile of $\chi_{\text{grid}}$ values in within 1000-km of the tropical cyclone center. This parameterization was shown to lead to skillful forecasts of tropical cyclone intensity. Here, we motivate the entropy deficit parameterization in this model with that used in \citeA{lin2020forecasts}, by computing $\chi$ as:

\begin{equation}
    \chi = \exp \big( \text{log} \: \chi_{\text{grid}} + \chi_\sigma \big) + \chi_a
\end{equation}
Since $\chi$ is approximately log-normal distributed, as in~\citeA{tang2012ventilation}, we add $\chi_\sigma$ to the logarithm of the monthly-mean gridded entropy deficit, $\chi_{\text{grid}}$, as well as $\chi_a$ to $\chi$ everywhere. In this study, we assume $\chi_\sigma$ and $\chi_a$ to be constant throughout all months, though future work could try to determine if this is choice is indeed optimal.

\section{Model Benchmarks}
For the purposes of this model development paper, we benchmark the model using a variety of comparisons to observations. Our comparisons of genesis, track, and intensity statistics are carried out on the global scale. We downscale ERA5 reanalysis data from 1979 to 2021, using monthly-averaged daily winds at 250- and 850-hPa, monthly-mean temperature and relative humidity, and monthly-mean sea surface temperature. Potential intensity is calculated using the new algorithm, under pseudoadiabatic lifting. There are some differences in the ensuing results when using the new potential intensity algorithm, as opposed to the original BE02 algorithm, but the results are not statistically robust. A total of $\approx 600,000$ tropical cyclone tracks are generated with the downscaling model, such that the sample sets for the resulting analysis are statistically robust. As emphasized in \citeA{emanuel2022tropical}, the model's tropical cyclone frequency must be normalized by a single constant. Here, we normalize the seeding rate such that the global annual TC count between 1979-2021 is equivalent to that of the observations.

On a modern laptop, we are able to produce around 1000 tracks per core-hour. The sample set used in this study can be generated in around 30 hours using a single CPU with 16-cores. Where applicable, the results are stratified by basin, as defined in the IBTrACS dataset \cite{knapp2010international}, except for the Southern Hemisphere basins, which are split into the South Indian ($30^{\circ}E-100^{\circ}E$), Australian ($100^{\circ}E-180^{\circ}E$), and South Pacific ($180^{\circ}E-260^{\circ}E$) basins.

\subsection{Genesis Statistics}

\begin{figure}
    \noindent\includegraphics[width=1.0\textwidth]{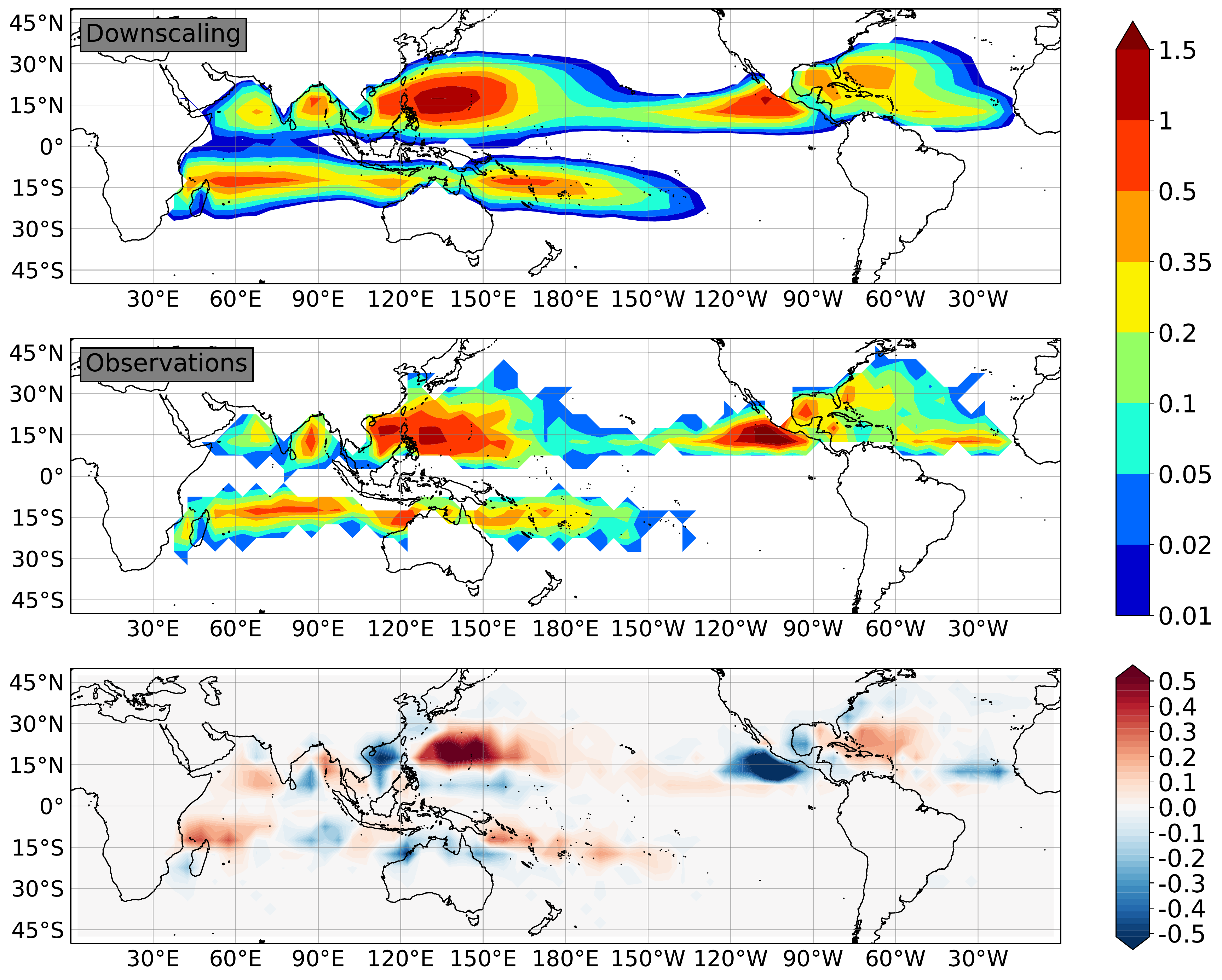}
    \caption{Number of genesis events per year, from (top) the downscaling model, and (middle) observations from 1979 to 2021. (Bottom) Difference in the number of genesis events per year (top minus middle), on a linear scale. $5^{\circ}$ by $5^{\circ}$ boxes are used to bin genesis events.}
    \label{fig:genesis_pdf}
\end{figure}

To begin, we compare the annual density of genesis events to historical observations from 1979 to 2021, using $5^{\circ}$ by $5^{\circ}$ boxes to bin events. Figure \ref{fig:genesis_pdf} shows that in general, the observed tropical cyclone genesis distribution is well simulated by the random seeding method combined with the FAST tropical cyclone intensity simulator \cite{emanuel2008hurricanes, emanuel2017fast}. However, there are a few biases in the model. For instance, the region of greatest probability of genesis seems to be biased too far eastward in the Western Pacific region, and genesis density is biased too low in the Eastern Pacific region. Furthermore, in the observational record, there is a regional peak in genesis rate right off the north-west coast of Australia, and in the Gulf of Carpentaria. Neither is immediately obvious in the genesis distribution of the downscaling model. The downscaling model's genesis rate in the Australian region is also slightly too high in the Southern Pacific. This is characteristic of genesis that is biased too far eastward in the South Pacific. A detailed comparison of the fraction of global tropical cyclone in each basin is shown in Figure S2. In addition, the downscaling model typically under-predicts genesis in extratropical regions. This bias could have several possible causes. First, monthly-mean moisture is used to drive the model. Moisture anomalies on time scales shorter than a month may be important to capture tropical cyclones in these regions, as they may temporarily elevate the genesis potential. These biases could also arise from the fact that the downscaling model's physics do not explicitly account for any kinetic energy derived from baroclinic instability \cite{davis2004tt}. Regardless of these biases, the major tropical cyclone genesis regions are well represented in the downscaling model.

\begin{figure}
    \noindent\includegraphics[width=1.0\textwidth]{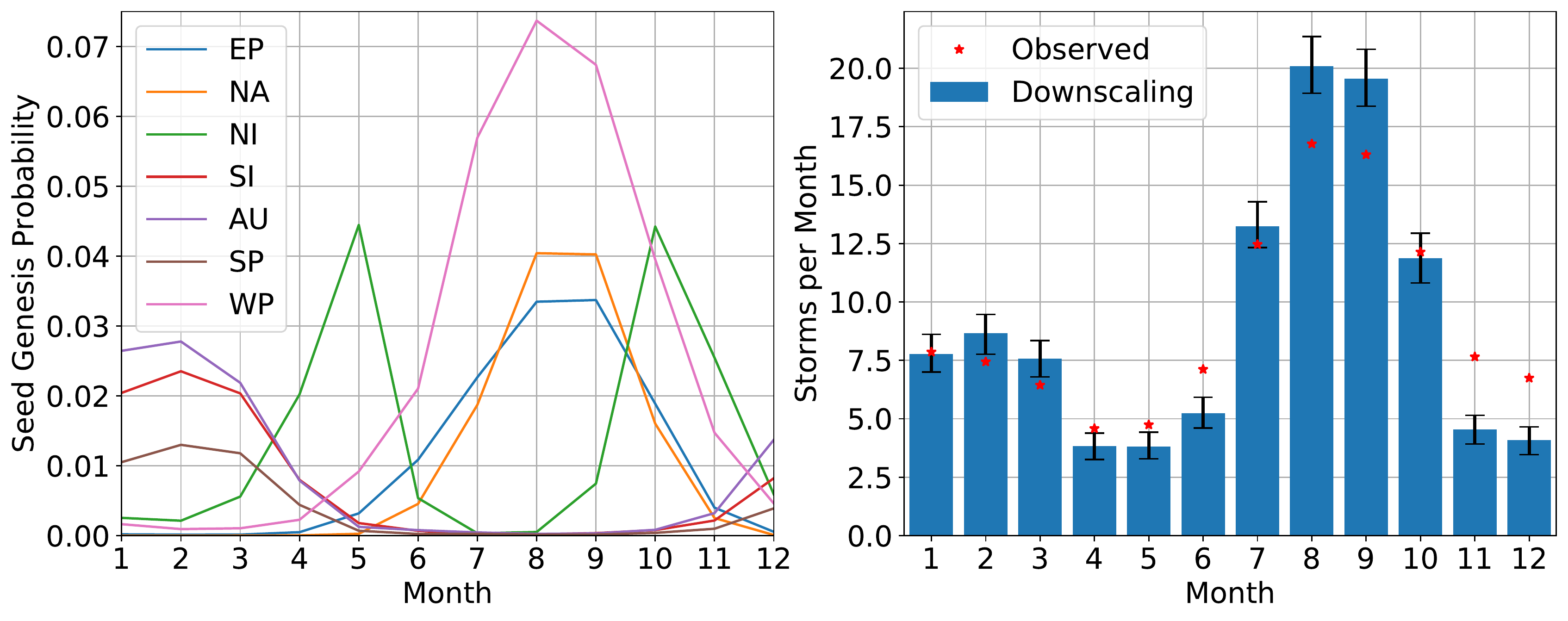}
    \caption{(Left): Probability that a weak seed will undergo tropical cyclone genesis, for each basin. (Right): Comparison of number of tropical cyclones per month between observations and the downscaling model. The downscaling model is normalized such that it has the same annual number of tropical cyclones as the observations. Error bars indicate the 95\% confidence interval when sub-sampling downscaling events to the same size as observational events.}
    \label{fig:seasonal_cycle}
\end{figure}

We also investigate the seasonal cycle of tropical cyclone genesis. First, we compute the seed genesis probability, or the chance that a weak seed will undergo tropical cyclone genesis (a maximum wind of greater than 18~m~s$^{-1}$). In the random seeding approach, a large number of the randomly placed seeds die and are thrown out \cite{emanuel2008hurricanes, emanuel2022tropical}, which is reflected in the low seed probabilities shown in Figure \ref{fig:seasonal_cycle}, left; globally, only around 1 in every 125 randomly placed seeds survives. Since the seeds are also placed randomly in time, the seed genesis probability shown in Figure \ref{fig:seasonal_cycle} also reflects the seasonal cycle of each individual tropical cyclone basin. In general, the downscaling model can represent the sharp tropical cyclone seasonal cycle in each basin, though there are slight negative biases in off-peak tropical cyclone months (for instance, May and November in the Atlantic basin). As mentioned earlier, consideration of moisture anomalies on time scales shorter than a month and/or inclusion of baroclinicity into the model physics could alleviate this bias. However, the off-peak tropical cyclones are typically weak, short-lived, and derivatives of baroclinic instabilities, and thus do not contribute strongly to the power dissipation index or the heavy-tail of tropical cyclone hazard.

The global seasonal cycle of tropical cyclone count is also reasonably represented in the downscaling model. Figure \ref{fig:seasonal_cycle}, right, shows the global seasonal cycle in tropical cyclone count, with error bars indicating the 95\% confidence interval when sub-sampling the downscaling events to be the same number as the historical record. Here, it is clear that the downscaling model underpredicts tropical cyclone count during the off-peak months in both basins, namely May to June and November through December. Since the downscaling events are normalized to have the same number of events as the historical period, tropical cyclone count is over-predicted during peak tropical cyclone months and under-predicted during off-peak tropical cyclone months. Nevertheless, the key components of the global seasonal cycle are well reproduced using the downscaling model.

Finally, we investigate the latitudinal distribution of genesis, using $3^{\circ}$ latitude bins. Again, we including the 95\% confidence interval from sub-sampling the downscaling events to the same size as the observational record. In general, the downscaling model faithfully represents the latitudinal distribution of genesis, though it underestimates tropical cyclone genesis in the extratropics, as discussed earlier.

\begin{figure}
    \noindent\includegraphics[width=0.7\textwidth]{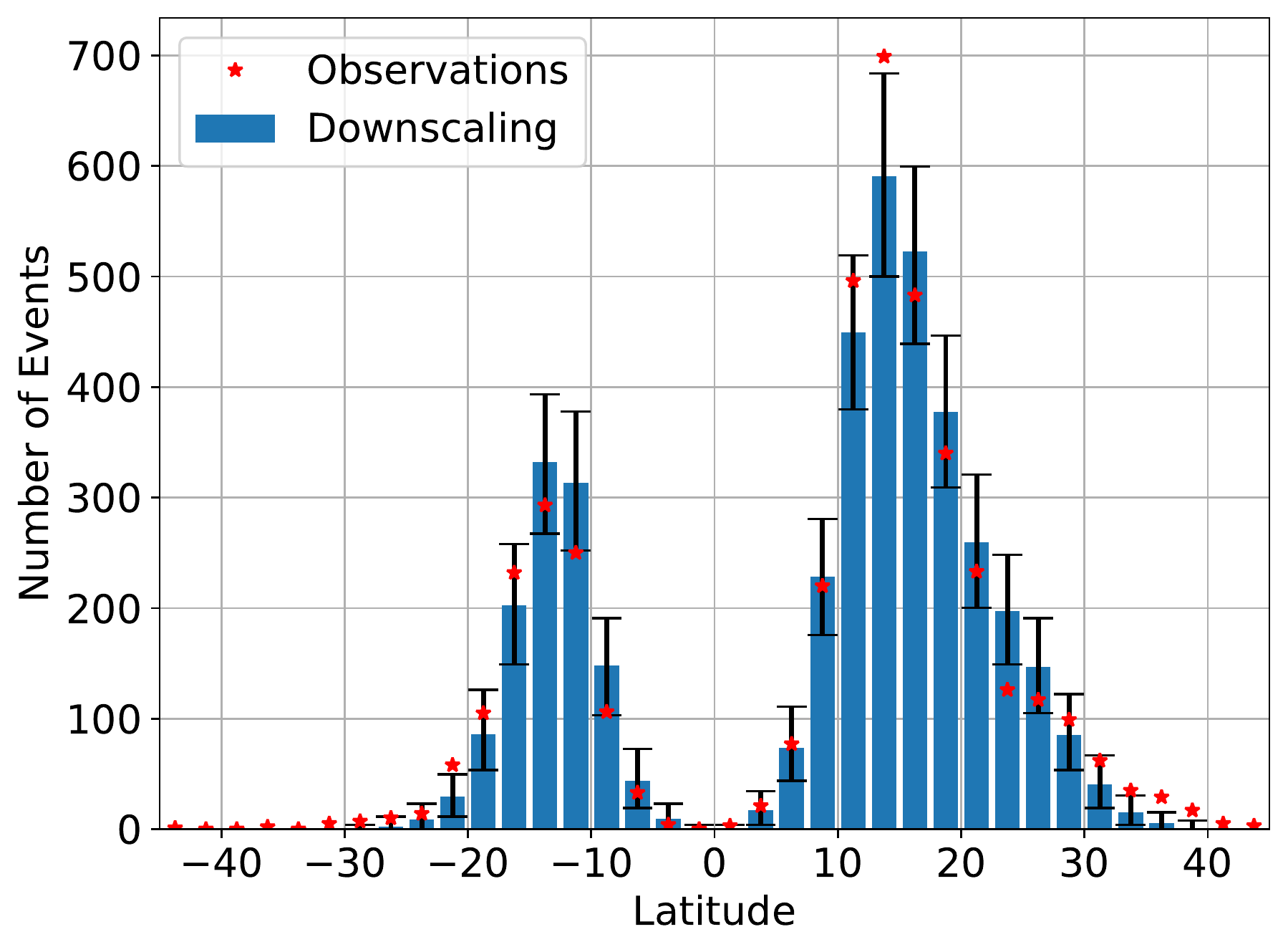}
    \caption{Distribution of the latitude of genesis. The downscaling distribution is normalized to have the same number of total genesis events as the observations. Error bars indicate the 95\% confidence interval when sub-sampling downscaling events to the same size as observational events.}
    \label{fig:AL_seasonal_cycle}
\end{figure}

\subsection{Track and Intensity Statistics}
\begin{figure}
    \noindent\includegraphics[width=1.0\textwidth]{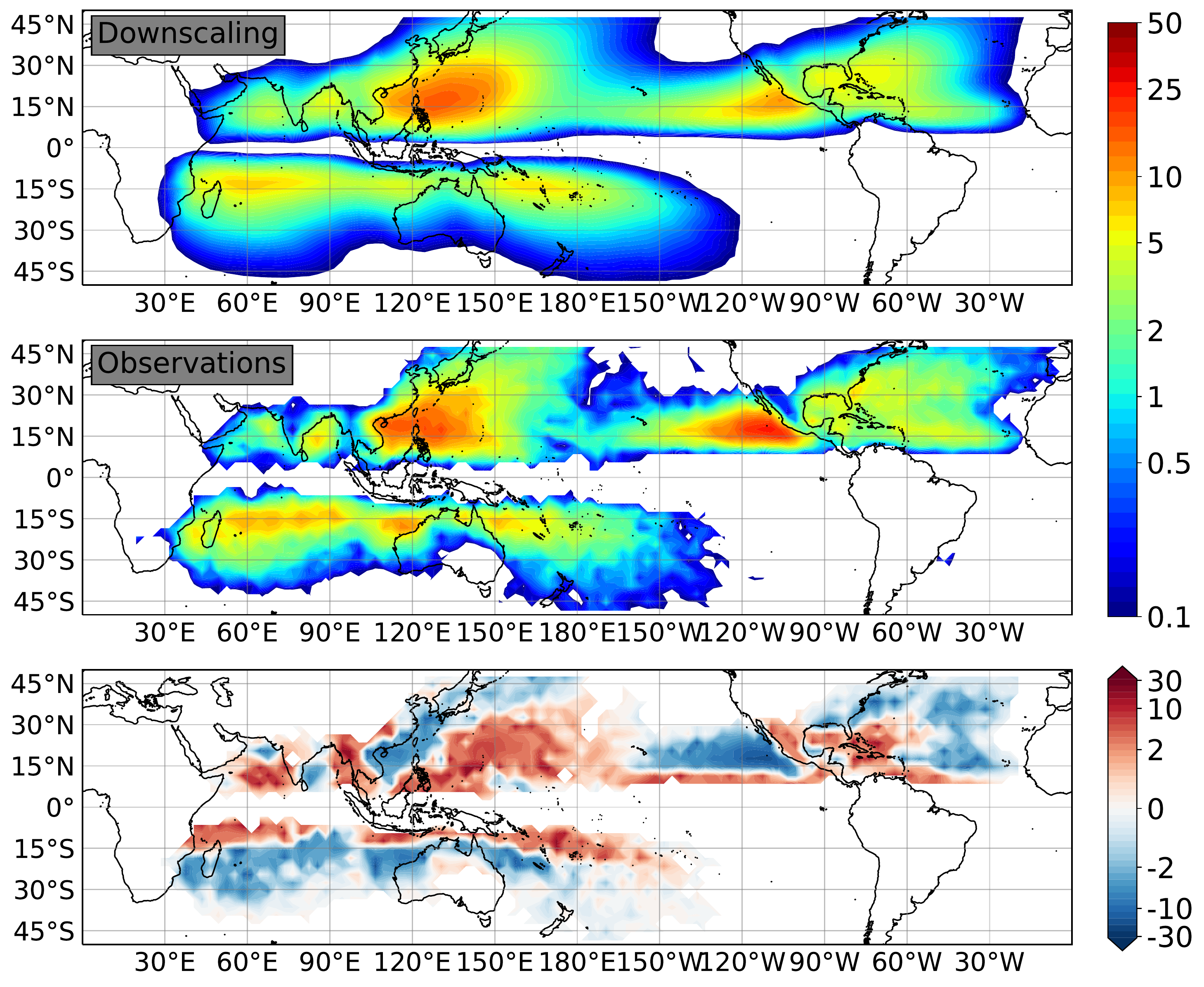}
    \caption{Number of 3-hourly track crossings per year, using $3^{\circ}$ by $3^{\circ}$ longitude-latitude boxes, from (top) the downscaling model, and (middle) observations from 1979 to 2021. (Bottom) Difference in number of 3-hourly track crossings per year, between the downscaling model and observations. The color scale is linear from -2 to 2, and logarithmic where the magnitude of the difference is greater than 2.}
    \label{fig:track_density}
\end{figure}

In this section, we analyze the track and intensity statistics of the tropical cyclones represented in the downscaling model. First, we look at number of 3-hourly track crossings per year, using $3^{\circ}$ by $3^{\circ}$ longitude-latitude boxes. The number of tracks in the downscaling model are normalized such that it has the same number of tropical cyclones per year as the observations. As shown in Figure \ref{fig:track_density}, the modeled track density distribution qualitatively represents that of the observational record, though there are a few notable biases. The bias of largest magnitude is the negative bias in track density over the Eastern Pacific region, which is most likely attributed to the negative bias in genesis in eastern portion of that region. In the Western Pacific region, the number of track crossings are of comparable magnitude between the model and the observations, though the general track of tropical cyclones are biased too far eastward in the downscaling model. There is also a negative bias in track density polewards of around $30^{\circ}N$ and $15^{\circ}S$ that again, could be alleviated through inclusion of baroclinic instability and/or moisture anomalies on time scales shorter than a month into the model physics. In general, it is hard to see differences in track density between the intensity-dependent $\alpha$ experiments and constant $\alpha$ experiments, since intense storms, which are most affected by intensity-dependent steering, comprise only a small fraction of the total tropical cyclone count.

\begin{figure}
    \noindent\includegraphics[width=0.7\textwidth]{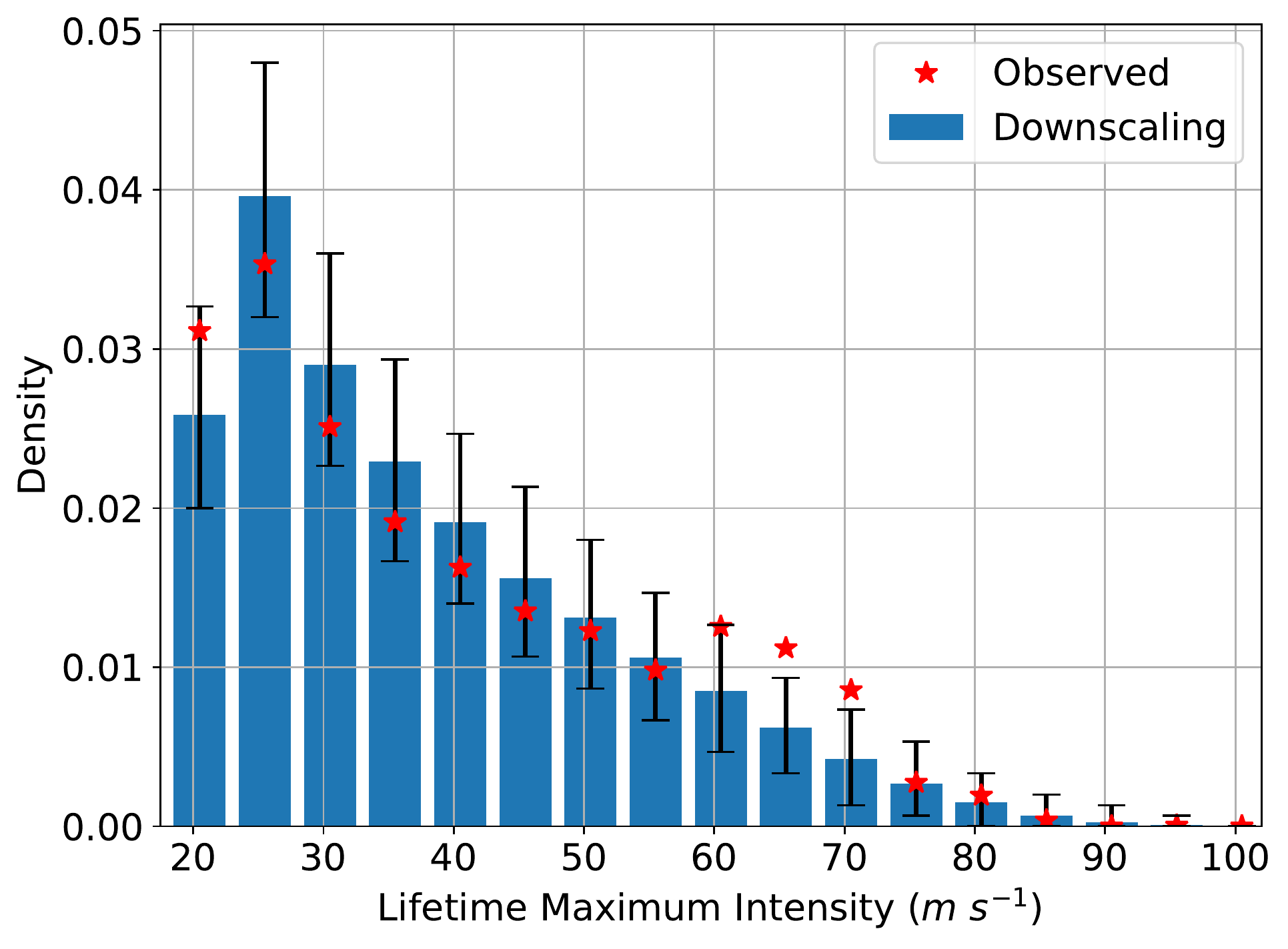}
    \caption{Comparison of the lifetime maximum intensity probability density distribution between the downscaling model and observational record, using 5~m~s$^{-1}$ wide bins. Error bars indicate the 95\% confidence interval when sub-sampling downscaling events to the same size as observational events.}
    \label{fig:lmi_distribution}
\end{figure}

We also show the distribution of the lifetime maximum intensity (LMI) of downscaled tropical cyclones. Figure \ref{fig:lmi_distribution} shows that the modeled lifetime maximum intensity distribution closely follows the observed distribution, with a peak around 25~m~s$^{-1}$ and an exponential decay in probability with increasing LMI. Here, it is important to note that the differences between the modeled and observational distributions are not statistically significant, except for the bi-modality in the distribution that is a direct result of rapidly intensifying storms \cite{lee2016rapid}. We do not make an explicit attempt to account for the bi-modality in the LMI distribution, and leave that for future work. We further note that the good performance of the model in reproducing the distribution of tropical cyclone lifetime maximum intensities warrants the use of the simplified rapid algorithm to compute $V_p$.

\subsection{Inter-annual Variability}
Finally, we investigate inter-annual variability in the downscaling model, by analyzing the downscaling model's ability to capture inter-annual variability in tropical cyclone activity. In the ensuing analysis, TCs that occur during austral summer are aggregated into the year at the end of austral summer. For example, TCs that occur from September 2000 to June 2001 in the Southern Hemisphere are considered to occur in 2001. This is done so that each data point represents Southern Hemisphere TCs that occur in the same TC season.

\begin{figure}
    \noindent\includegraphics[width=1.0\textwidth]{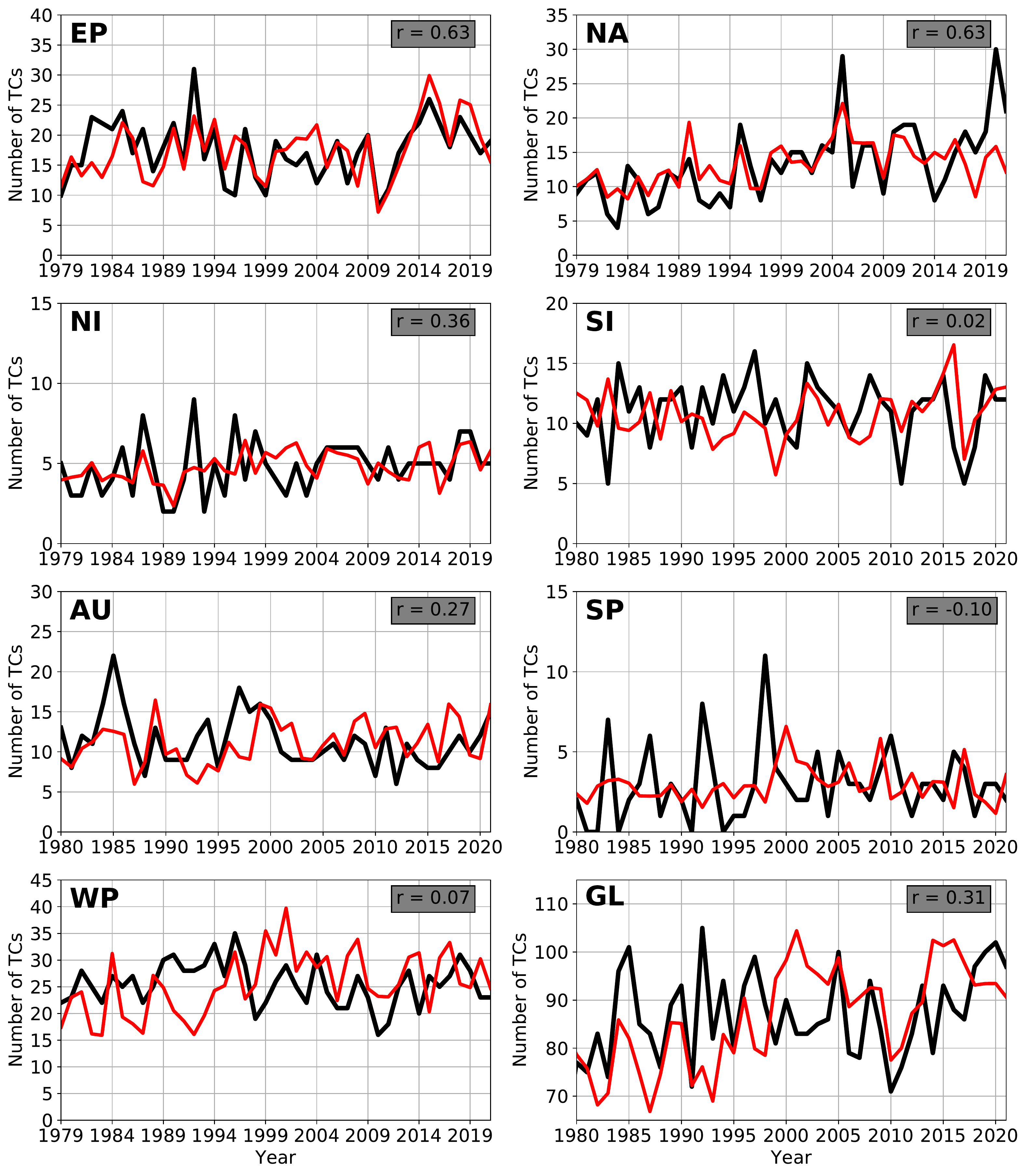}
    \caption{Inter-annual variability in the number of tropical cyclones for each basin, from the (black) observational record and the (red) downscaling model. Basin tropical cyclone counts in the downscaling model are normalized by the average tropical cyclone count over the historical period in each basin. Pearson correlation coefficients are shown in the top-right of each panel. Only storms where the LMI is greater than 18~m~s$^{-1}$ are considered.}
    \label{fig:interannual_tc}
\end{figure}

In the random seeding approach, seeds are randomly placed in space and time at a constant rate, such that inter-annual variability in tropical cyclone count is also a measure of inter-annual variations in probability that a weak seed intensifies into a tropical cyclone. Thus, in this model, inter-annual variability comes from inter-annual changes in the large-scale environment, which ultimately determines the transition probability of the weak proto-vortex into a tropical cyclone. Figure \ref{fig:interannual_tc} shows that the downscaling model is also able to reasonably capture inter-annual variability in tropical cyclone count, particularly in the Eastern Pacific and North Atlantic regions, where genesis potential indices have high skill \cite{camargo2007use}. The values of the correlation coefficients are comparable, if not higher, than those shown in \cite{lee2018environmentally}, though the years analyzed in that study were from 1981-2012. There is very little correlation in inter-annual variability in the West Pacific basin, which is a documented deficiency of genesis potential indices \cite{menkes2012comparison}. ENSO, the main source of tropical interannual variability, has single-signed signals in tropical cyclone genesis in the Atlantic and Eastern Pacific basins, but mixed-signed signals in the Western Pacific, South Pacific, and South Indian basins [see Supporting Information, Figures S3 and S4, and \citeA{camargo2007use}]. It is likely that the interannual signal in genesis gets averaged out in basins such as the Western Pacific. Though this model is not directly based on a genesis potential index, it uses similar input variables. Finally there is also decent correlation of inter-annual global tropical cyclone count ($r = 0.31$), mostly owing to high inter-annual skill in the Eastern Pacific and North Atlantic regions. 

\begin{figure}
    \noindent\includegraphics[width=1.0\textwidth]{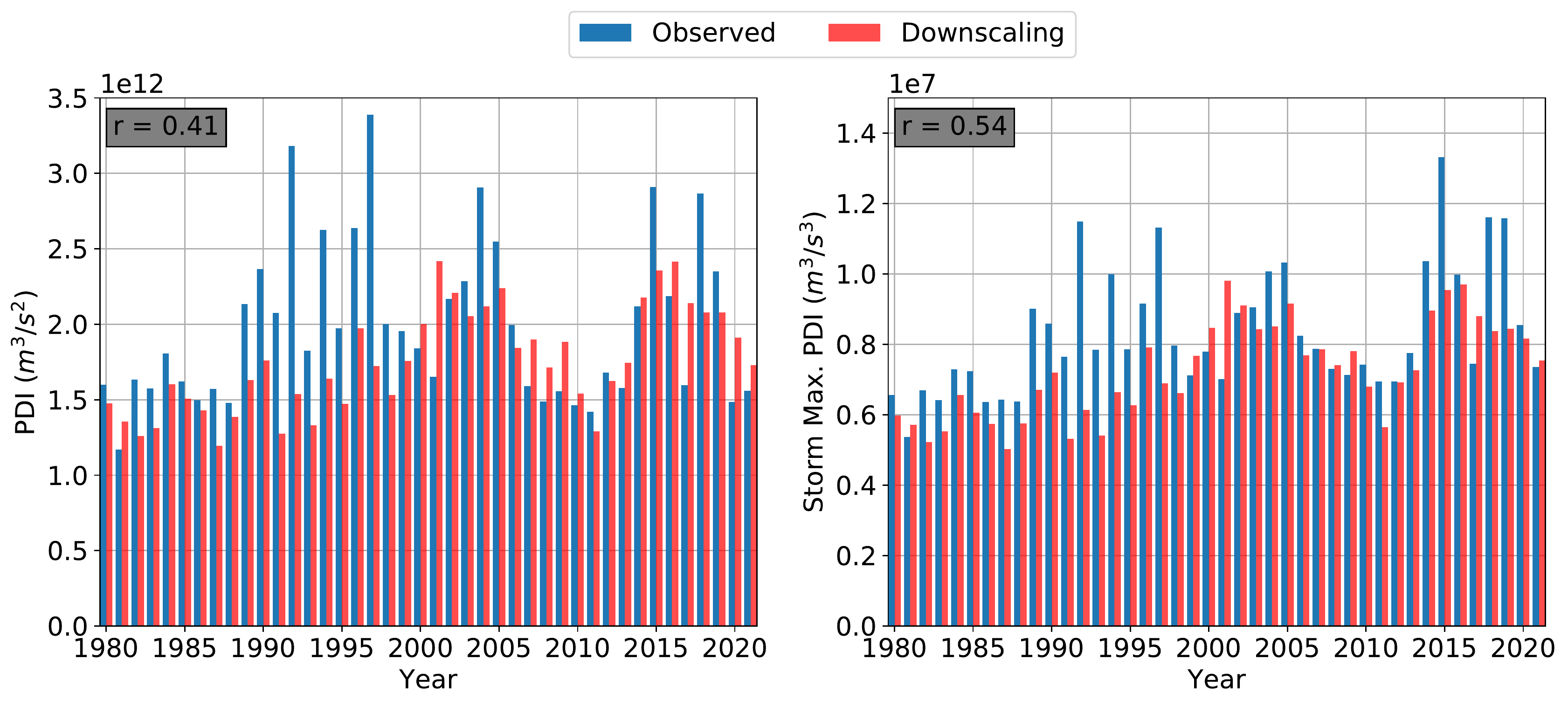}
    \caption{Inter-annual variability in the global (left) power dissipation index and the (right) storm maximum power dissipation index. Only storms where the LMI is greater than 18 m~s~$^{-1}$ are considered.}
    \label{fig:interannual_pdi}
\end{figure}

Another metric that is arguably more predictable (or less noisy) than the global tropical cyclone count is the power dissipation index (PDI). The PDI is calculated as integral of the cube of the storm intensity over its entire lifetime, over all tropical cyclones in a year. Thus, PDI accounts for not only tropical cyclone frequency, but also duration and intensity. Figure \ref{fig:interannual_pdi} compares historical inter-annual variations in the global PDI with that predicted by the downscaling model. The correlation coefficient is $r = 0.41$, showing that the downscaling model is also able to decently capture global inter-annual variations in the PDI. The PDI correlation is strongly influenced by outliers in the 1990s; the correlation increases to $r = 0.61$ when sub-setting the historical period to years after 2000s. We also calculate the storm maximum PDI, which is a simplified version of PDI and is calculated as the sum of the cube of the storm lifetime maximum intensity, over all tropical cyclones in a year. In this sense, storm maximum PDI does not include the overall lifetime of the tropical cyclone. The correlation coefficient is $r = 0.54$, indicating that model skill improves when consider only storm frequency and maximum intensity. Note, we do not bias correct the annual-average of the downscaled global PDI, which is around 90\% the annual-averaged global PDI in the observations. Since the global frequency in the downscaled model is normalized to be the same as that of the historical observations, this bias is a result of modeling biases in storm duration and intensity. Defining genesis as the first time point when the TC reaches 18~m~s$^{-1}$, and lysis as the last time point the TC exceeded 18~m~s$^{-1}$, then the mean storm lifetime in this downscaling model is around 5.3 days, whereas the mean storm lifetime in the observations is around 4.2 days. Note that these calculations are sensitive to the intensity threshold used. However, since PDI is also weighted by the cube of intensity, biases in the frequency of intense storms, compared to observations, have a much stronger impact on the modeled PDI. As implied in Figure \ref{fig:lmi_distribution}, the downscaling model has a negative bias in the frequency of intense storms (the reasons for which this occurs were discussed in the previous section), which is largely responsible for the negative bias in global PDI.

\section{Tropical Cyclone Hazard}
Finally, in this section, we will consider global tropical cyclone hazard, which combines information about the genesis, track, and intensity evolution of tropical cyclones. Here, we consider the return period of tropical cyclones that have an intensity of at least 33~m~s$^{-1}$ (Category 1 status). We calculate return period using $1^{\circ}$ by $1^{\circ}$ longitude-latitude boxes using both the observational data and downscaling events, ensuring not to double count singular events. Since the sample size of the downscaling events is much larger than that of the historical data, we use a Gaussian kernel of unit standard deviation to smooth the observational counts. The return period, as calculated, is thus defined as how often a $1^{\circ}$ by $1^{\circ}$ grid-box will observe tropical cyclone of intensity of at least 33~m~s$^{-1}$.

\begin{figure}
    \noindent\includegraphics[width=1.0\textwidth]{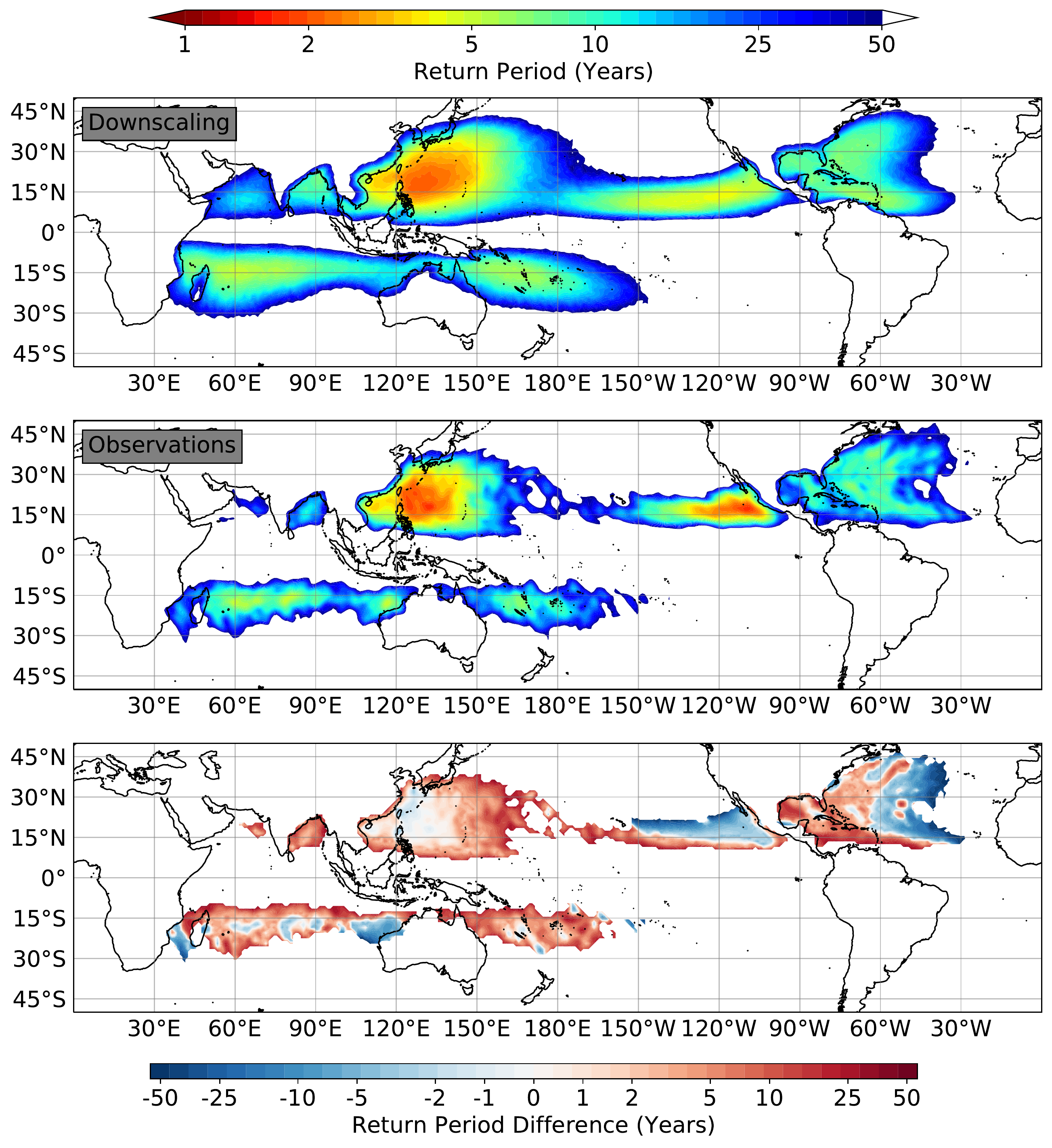}
    \caption{Global map of the return period of tropical cyclones that reach an intensity of at least 33~m~s$^{-1}$, from the (top) downscaling model and from (middle) observations, using $1^{\circ}$ by $1^{\circ}$ longitude-latitude boxes. A Gaussian kernel of unit standard deviation is used to smooth the observations. (Bottom) The difference in return period between downscaling and observations (panel (b) minus panel (a)). Blue (red) shading is where hazard is underestimated (overestimated). The color scale is linear from -2 to 2 years, and logarithmic where the magnitude of the return period difference is greater than 2 years.}
    \label{fig:return_period_cat1}
\end{figure}

Figure \ref{fig:return_period_cat1} compares the calculated return period of tropical cyclones that have an intensity of at least 33~m~s$^{-1}$ between the downscaling events and the observational record. Here, we consider hazard to be overestimated when, for a fixed a hazard, the return period in the observations is larger than that estimated by the downscaling model. Note that at the interfaces between areas where tropical cyclones are observed and those where there is no tropical cyclone activity, the downscaling model will tend to overestimate the return period (underestimate hazard), since the sample size of the observational record is much smaller than those of the downscaling model. There is generally very little disagreement in return period in the Western Pacific basin, whereas there seems to be a southward bias in the region of smallest return periods (i.e. where a hurricane is most likely) in the Eastern Pacific basin. The magnitude of return period differences are generally not large in the Atlantic basin either (around 2 years in magnitude on average), except for the western Gulf of Mexico region, where the downscaling model seems to overestimate hazard with respect to the historical record, though the historical record has larger uncertainties at longer return periods.

\begin{figure}
    \noindent\includegraphics[width=1.0\textwidth]{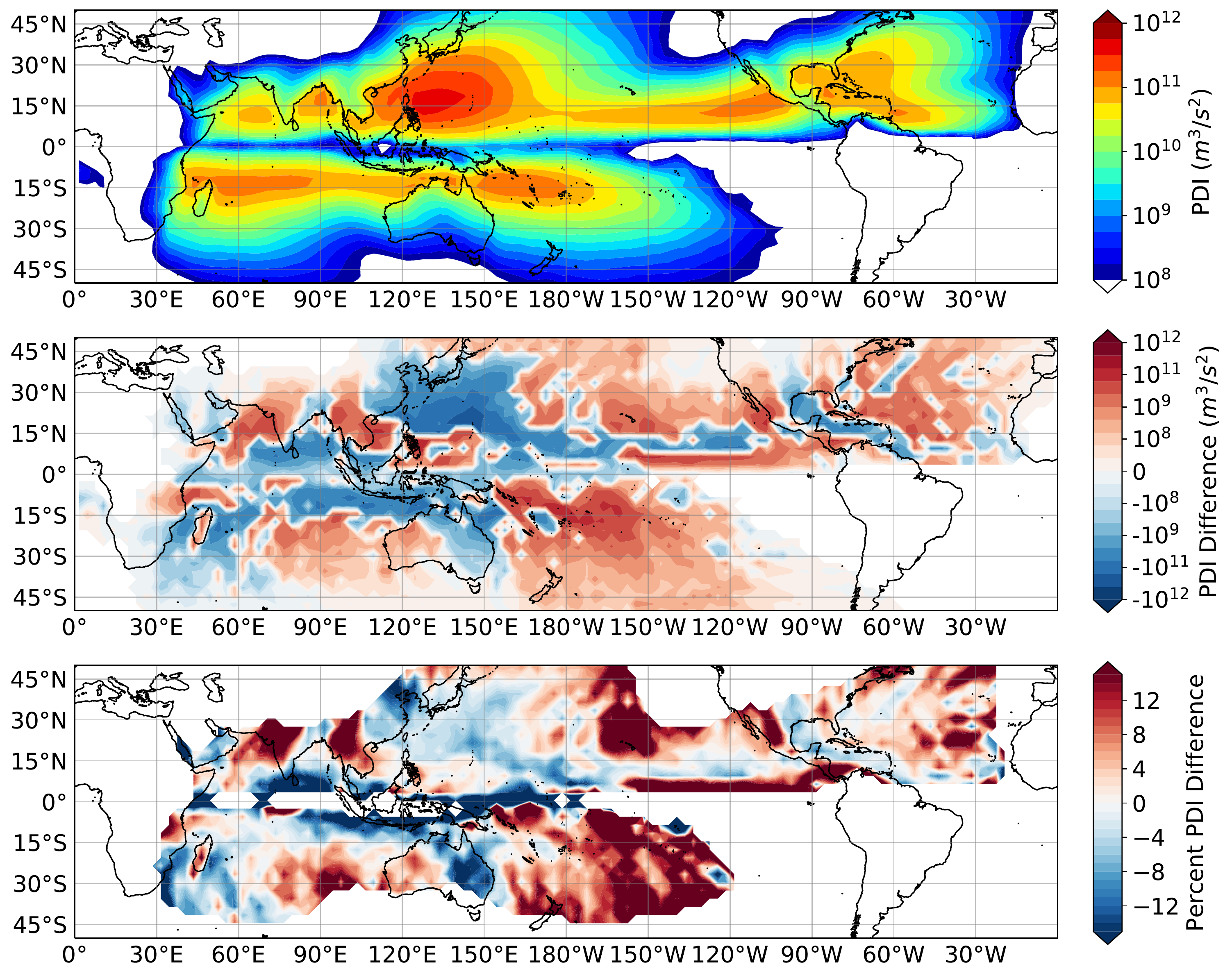}
    \caption{(Top) Mean PDI in the control downscaling experiment over the 43-year reanalysis period, using 3$^\circ$ by 3$^\circ$ bins. (Middle) Difference in the mean PDI between the intensity dependent $\alpha$ and the constant $\alpha$ simulations. The scale is linear from -$10^8$ to $10^8$, and logarithmic for differences with magnitude above $10^8$. Red (blue) shading indicates more (less) TC activity in the variable-alpha experiments than in the control. (Bottom) Percent difference in PDI from the intensity dependent $\alpha$ to the constant $\alpha$ simulations, where grid-points with a mean PDI less than $10^8$ are removed.}
    \label{fig:pdi_map}
\end{figure}

It is also worth commenting on how the intensity-dependent $\alpha$ changes the general distribution of major tropical cyclone activity. Since $\alpha$ has the largest differences at the strongest of intensities, we use PDI to understand how an intensity-dependent $\alpha$ influences major tropical cyclone activity. Figure \ref{fig:pdi_map} shows the mean PDI in the downscaling model, as well as differences in PDI between the intensity dependent $\alpha$ and constant $\alpha$ experiments. In general, the intensity-dependent alpha expands the region of TC activity - the increases are, for the most part, at the margins of the regions of greatest PDI in the control simulation, while the core TC regions see decreases. However, there is also large regional variability in how PDI changes. For instance, PDI decreases in the Caribbean Sea and Western Gulf of Mexico, while it generally increases over the North Atlantic Ocean. Furthermore, PDI increases in the South China Sea but decreases over the northern part of the sea. The latter can be directly attributed to the presence of mean-easterlies and southerlies at 250-hPa during boreal summer. In the intensity-dependent $\alpha$ experiments, the more intense a tropical cyclone, the more its track will follow the upper-level winds. In general, this is the observed pattern when considering differences between the intensity-dependent $\alpha$ and constant $\alpha$ experiments. Since there is considerable regional variability in the upper-level zonal and meridional winds, there is also much regional variability in the TC activity response when adding intensity-dependent steering. Investigation of percent changes to the PDI (Figure \ref{fig:pdi_map}, bottom), shows that in some regions, an intensity-dependent $\alpha$ can lead to a 5-10\% change in the PDI.

\begin{figure}
    \noindent\includegraphics[width=1.0\textwidth]{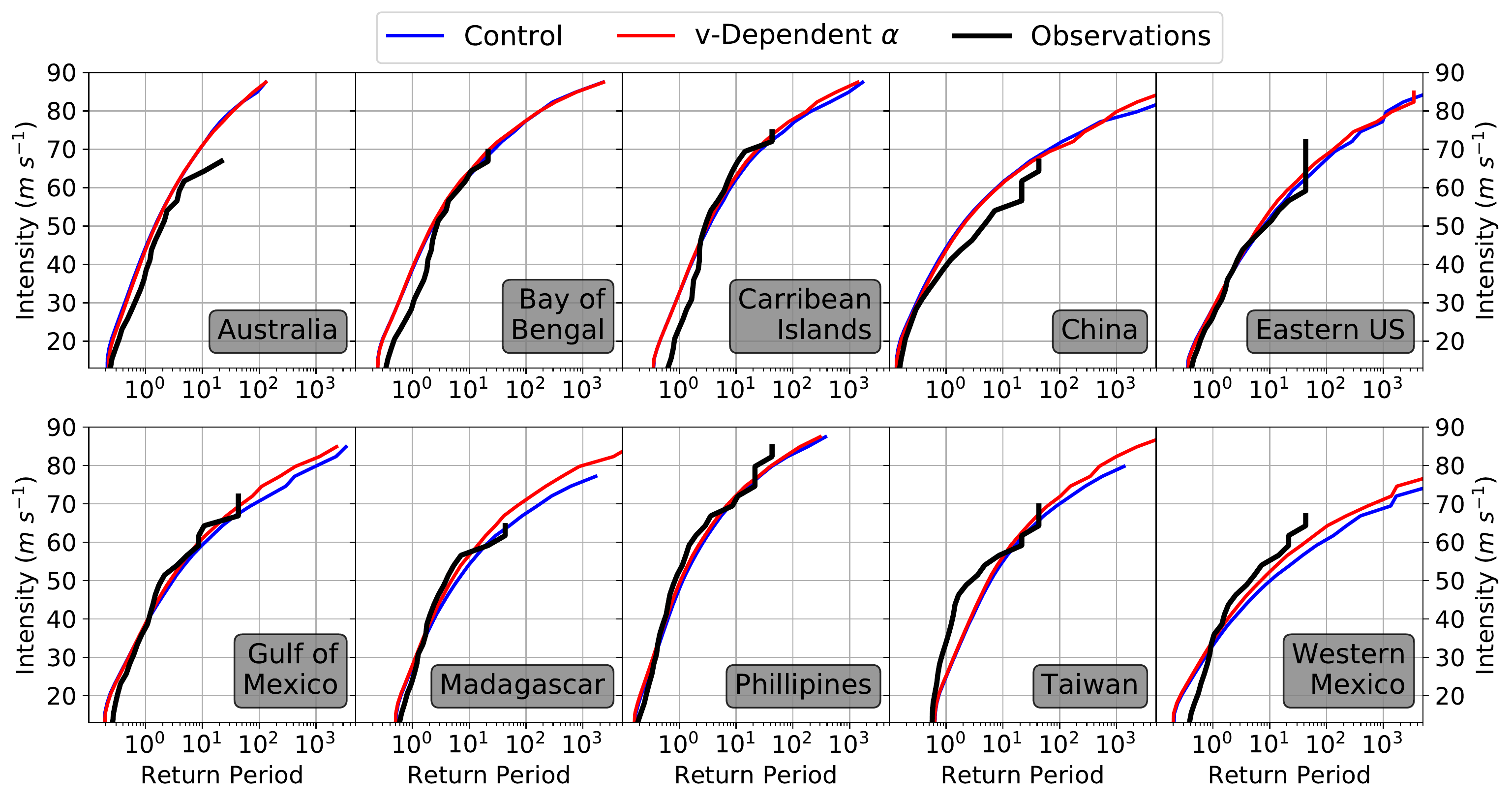}
    \caption{Return period curves of landfall intensity at labeled areas around the world, from all tracks in the (blue) control and (red) intensity-dependent $\alpha$ downscaling model. Return period curves calculated from observations are in black.}
    \label{fig:return_period}
\end{figure}

Finally, we calculate return period curves of landfall intensity at various areas around the globe that are prone to tropical cyclones. Return period curves are valuable since they highlight the frequency of the strongest of tropical cyclones, which often are the most destructive and costly. Each region is defined following \citeA{lee2018environmentally}, finding all locations over land that are within 50-km of a coastline. Figure \ref{fig:return_period} shows return period curves of landfall intensity at various regional locations, calculated from the control and intensity-dependent $\alpha$ downscaling experiments. The return period curves are benchmarked against return periods estimated from observational data, and are not bias-corrected to the observations.

In general, we observe that the return period curves are in agreement with those derived from observations at low intensities, though there are small biases, such as an overestimation of the return period of weak storms in Western Mexico, the Bay of Bengal, and the Caribbean Islands. It is also informative to analyze the difference in return period curves between the control and intensity-dependent $\alpha$ downscaling experiments, since return period curves magnify the tail of the tropical cyclone distribution. In particular, the return period curves show that the frequency of the most intense storms increases along the Gulf of Mexico, Madagascar, Western Mexico coastlines, while there are no discernible differences in Australia, the Bay of Bengal, the Caribbean Islands, the Philippines, and China. Note that for some regions, such as the Northern Australia Coast and the Gulf of Mexico, one portion of the coastline sees an increased frequency of major tropical cyclones under the intensity-dependent $\alpha$ model, while other portions of the same coastline see a decreased frequency of major tropical cyclones. Whether or not differences between these return period curves will increase or decrease with warming is an important and interesting question, and one that will be the subject of future work.

\section{Summary and Discussion}
In this study, we develop an open-source, physics-based tropical cyclone downscaling model. The model synthesizes concepts from the MIT tropical cyclone downscaling model \cite{emanuel2004environmental, emanuel2008hurricanes, emanuel2022tropical}, randomly seeding weak vortices in space and time and evolving them within the large-scale environment. The weak seeds translate according to the beta-and-advection model \cite{marks1992beta}, and intensify according to the FAST intensity model \cite{emanuel2017fast}. Only seeds that reach traditionally defined tropical storm strength are kept. A number of changes are made to the MIT tropical cyclone downscaling model. In particular, we include a dependence of the depth of the steering flow on intensity, introduce a new Python-based algorithm to calculate potential intensity, incorporate a new parameterization of ventilation in the FAST intensity model, and expand the same intensity model to the global scale.

Using these methods, the model is shown to reasonably represent the climatology of tropical cyclone activity, as compared to the observational record. A number of benchmarks are used evaluate the model. We show that the tropical cyclone downscaling model's seasonal cycle, genesis, track density, and intensity distributions are generally close to the observational record, though there are a few biases as discussed in the main text. Furthermore, correlations in inter-annual tropical cyclone count are comparable to those of genesis potential indices \cite{camargo2007use}. The downscaling model also displays substantial correlation with the historical record of global storm maximum power dissipation index. We also compared return periods of storms that reach an intensity of at-least 33~m~s$^{-1}$, and found general agreement between return periods calculated from the downscaling model and those calculated from historical data.

The genesis method is based on random seeding, as opposed to a statistically trained algorithms that directly reproduces observed tropical cyclone genesis patterns. This should be seen as both a strength and a weakness of this model. For instance, while there are a few biases in the genesis patterns, as shown in Figure \ref{fig:genesis_pdf}, the genesis pattern does not depend on the sparse sampling set over the historical period. Research has also shown that tropical cyclone frequency, as predicted by downscaling models, can rapidly diverge in future warming scenarios, depending on whether relative humidity or saturation entropy deficit is used in statistical indices of tropical cyclone genesis \cite{lee2020statistical}. This is because both quantities vary in synchronicity in the current climate, but diverge in warming scenarios. The random seeding approach does not resolve that issue, but rather, presents an alternative approach, as discussed thoroughly by \citeA{emanuel2022tropical}. However, it is worth highlighting this model's dependence on both quantities, as relative humidity plays a role in initializing the inner core moisture of the intensity model, while the saturation entropy deficit modulates the rate at which the inner core moisture dries through ventilation. Future changes to both variables would play a role in the genesis rate predicted by this model.

Furthermore, while the parameterization of ventilation in the intensity component of the downscaling model has been evaluated in the same intensity model on forecasting time scales \cite{lin2020forecasts}, the success of this parameterization in a forecasting model by no means guarantees its correctness in its response to warming. This is primarily because the temperature dependence of the parameterization cannot easily be tested in the current climate since temperature fluctuations in the tropics are weak \cite{sobel2001weak}. The ventilation process, however, has support from theory and idealized numerical modeling, though it was primarily tested in mature tropical cyclones \cite{tang2010midlevel}. Recent work has additionally suggested that ventilation seems to play a large role in modulating tropical cyclone frequency under warming scenarios in numerical models \cite{hsieh2020large, hsieh2022model}. Still, an open question is whether or not ventilation (as opposed to some other variable) plays the dominant role in modulating the frequency and intensification rate of precursor tropical disturbances. In this model, the ventilation process has no intensity dependence, i.e. the randomly seeded proto-vortices and most intense of tropical cyclones are equally affected by the environmental saturation entropy deficit. How this assumption modulates this model's response to warming will be the subject of future research.

Despite these open problems, this physics-based downscaling model can be used to understand how physical processes in the large-scale environment play a role in modulating tropical cyclone genesis, track, and intensification. Because this model does not significantly depend on statistical sampling of historical tracks, it can, in principle, reproduce tropical cyclone variability in the climate system on decadal and multi-decadal time scales. This is one advantage of this model. The model can also be coupled with parametric models of tropical cyclone precipitation, as done in \citeA{lu2018assessing} and \citeA{feldmann2019estimation}. In addition, while we only presented results from downscaling reanalysis data, climate models can also be downscaled, though additional tuning and/or bias correction may be necessary. The behavior of tropical cyclones in different climates (and model representations of those climates) can be linked to specific processes in the atmosphere given the physical basis of the downscaling model. Furthermore, while the parameters [see Table \ref{table:parameters} for a summary] we used in this study lead to reasonable representations of tropical cyclone climatology, they should not be thought of as fixed. The model source code is freely available online for those interested in exploring the parameter space. Finally, the downscaling model may appeal to those interested in tropical cyclone hazard, since a large number of synthetic events can be rapidly generated.

\appendix
\section{Additional Model Information}
Table \ref{table:parameters} shows the summary of parameters used in the downscaling model. All of the variables are described in detail in the main text.

\begin{table}[h]
\caption{Summary of parameters used in the downscaling model. $\xi$ varies by basin and is shown in order for the Eastern Pacific, North Atlantic, North Indian, Western Pacific, Australia, South Pacific, and South Indian basins.}
\centering
\begin{tabular}{l c}
\hline
Variable & Value \\
\hline
$\phi_0$ & $2^{\circ}$ \\
$\xi$ & [6, 7, 2.5, 3.5, 6, 7, 3] \\
$v_{\text{init}}$ & 5 $\text{m s}^{-1}$ \\
$v_{\text{2d}}$ & 7 $\text{m s}^{-1}$ \\
$v_{\text{min}}$ & 15 $\text{m s}^{-1}$ \\
$v^*_{\text{min}}$ & 18 $\text{m s}^{-1}$ \\
$u_\beta$ & -1.0 $\text{m s}^{-1}$ \\
$v_\beta$ & 2.5 $\text{m s}^{-1}$ \\
$b_\alpha$ & 0.83 \\
$m_\alpha$ & 0.0013 $\text{(m/s)}^{-1}$ \\
$\alpha_{\text{min}}$ & 0.59 \\
$\alpha_{\text{max}}$ & 0.83 \\
$S_w$ & 0.80 \\
$\chi_\sigma$ & 0.5 \\
$\chi_a$ & 1.3 \\
\hline
\end{tabular}
\label{table:parameters}
\end{table}

As described in \citeA{lin2020forecasts}, Equations \ref{eq_unet} and \ref{eq_G} described the function that converts the axisymmetric wind to a maximum wind speed of the tropical cyclone.
\begin{equation}
    \textbf{v}_{\text{net}} = \textbf{v} + G \textbf{u}_t + 0.1 \textbf{S} \frac{v}{15}
    \label{eq_unet}
\end{equation}
\begin{equation}
    G = \min \Big[1, 0.8 + 0.35 \Big(1 + \tanh \Big( \frac{\phi - 35}{10} \Big)  \Big) \Big]
    \label{eq_G}
\end{equation}
where $v$ is the maximum axisymmetric wind (as predicted by the intensity model), $\textbf{v}$ is the vector of axisymmetric wind, $\textbf{u}_t$ is the vector of the tropical cyclone's translational speed, $\textbf{S}$ is the vector of the environmental vertical wind shear, and $\phi$ is the latitude of the storm center. The maximum wind speed, $v^*$, is determined by taking the maximum of $\textbf{v}_{\text{net}}$ over the domain.

%



%
%

\section*{Open Research Section}
The daily ERA5 data for zonal and meridional winds are available at \url{https://cds.climate.copernicus.eu/cdsapp#!/dataset/reanalysis-era5-pressure-levels} via DOI: 10.24381/cds.bd0915c6 \cite{hersbach2018era5_hourly}. The monthly-averaged ERA5 data for temperature and specific humidity are available at \url{https://cds.climate.copernicus.eu/cdsapp#!/dataset/reanalysis-era5-pressure-levels-monthly-means} via DOI: 10.24381/cds.6860a573 \cite{hersbach2019era5_pressure}. The monthly-mean ERA5 data for sea-surface temperature and surface pressure fields are available at \url{https://cds.climate.copernicus.eu/cdsapp#!/dataset/reanalysis-era5-single-levels-monthly-means} via DOI: 10.24381/cds.f17050d7 \cite{hersbach2019era5_single}. The ERA5 reanalysis data are accessible by creating an account with the Climate Data Store service, and usable according to ECMWF license to use Copernicus products. The IBTrACS data used for evaluation of the model with observations are available at \url{https://www.ncei.noaa.gov/products/international-best-track-archive} via DOI: 10.25921/82ty-9e16 \cite{knapp2018international}.

The physics-based tropical cyclone risk model is freely available at \url{https://github.com/linjonathan/tropical_cyclone_risk} \cite{jonathan_lin_2023_7651063}. Code to generate the data, as well as instructions to run the model, are all available at the aforementioned link.

\acknowledgments
J. Lin gratefully acknowledges the support of the National Science Foundation through the NSF-AGS Postdoctoral Fellowship, under award number AGS-PRF-2201441. C.-Y. Lee is supported by the Palisades Geophysical Institute (PGI) Young Scientist award from Lamont-Doherty Earth Observatory, Columbia University. C.-Y. Lee and A. Sobel also gratefully acknowledge support from the Swiss Re Foundation. The authors also thank two anonymous reviewers, as well as Kerry Emanuel for insightful comments on an earlier version of the manuscript.

%
%

\bibliography{references}

%
%
%
%
%

\end{document}